\documentclass[prd,preprint,showpacs]{revtex4}
\usepackage{amsmath}
\usepackage{latexsym}
\usepackage{amsfonts}
\usepackage{graphicx}
\newcommand{\be}{\begin{equation}}
\newcommand{\ee}{\end{equation}}
\begin{document}
\title{Fixed points in interacting dark energy models}
\author{Xi-ming Chen} \email{chenxm@cqupt.edu.cn}
\affiliation{College of Mathematics and Physics, Chongqing University of Posts and
Telecommunications, Chongqing 400065, China}
\author{Yungui Gong} \email{gongyg@cqupt.edu.cn}
\affiliation{College of Mathematics and Physics, Chongqing University of Posts and
Telecommunications, Chongqing 400065, China}

\begin{abstract}
The dynamical behaviors of two interacting dark energy models are considered. In addition
to the scaling attractors found in the non-interacting quintessence model with exponential
potential, new accelerated scaling attractors are also found in the interacting dark
energy models. The coincidence problem is reduced to the choice of parameters in the
interacting dark energy models.
\end{abstract}

\preprint{arXiv: 0811.1698}

\maketitle

\section{Introduction}

There exists mounting evidences that the Universe is experiencing
accelerated expansion, driven by an unknown energy component called
``dark energy". The nature and origin of dark energy have been an
active research topic in the past years. Because the only observable
effect of dark energy is manifested by gravitational interaction, we
know nothing about the nature of dark energy except that it has
negative pressure. One simple dark energy candidate which is
consistent with current observations is the cosmological constant.
However, the small value of the vacuum energy density imposes a big
challenge to particle physics. Furthermore, the cosmological
constant model faces the ``coincidence" problem: Why is the dark
matter energy density comparable to the dark energy density now?

To alleviate the coincidence problem, other dynamical dark energy models were proposed, such as
the quintessence model \cite{quint}, the holographic dark energy model \cite{holo}, the Chaplygin gas
model \cite{chaplyg}, and the tachyonic model \cite{tach}. Recently, the weak gravity conjecture was
used to constrain the property of dark energy \cite{weak}. It is also possible that the Einstein theory of
gravity needs to be modified in order to explain the accelerated expansion. These models include
the $1/R$ gravity \cite{invr}, the $f(R)$ gravity \cite{fr}, the DGP model \cite{dgp}, and
string inspired models \cite{brane}.

The attractor solution is independent of initial conditions. If the dark energy model has an
accelerated scaling attractor solution and the ratio of the energy density
$\Omega_{\rm dark\ energy}/\Omega_{\rm dark\ matter}$ between the two dark sectors is order 1,
then the coincidence problem can be alleviated. It is well known that for the quintessence
model, exponential potentials have scaling attractor solutions \cite{expon}. For more general
scalar field model, scaling attractor solutions were also obtained \cite{gong}. In this
Letter, we discuss the dynamical behaviors of the quintessence model with exponential
potential $V(\phi)=V_0\exp(-\kappa\lambda\phi)$, here $\kappa^2=8\pi G$.
Since the late-time accelerated scaling attractors of the exponential potential is
the scalar field dominated solution, $\Omega_{\rm dark\ energy}=1$ \cite{expon}, it does
not provide a satisfactory solution for the coincidence problem. In general, the
dark energy may not evolve independently, a coupling between the dark matter
and dark energy is possible \cite{intmod}. With the interaction between the dark sectors, accelerated
scaling attractors with $\Omega_{\rm dark\ energy}/\Omega_{\rm dark\ matter}=O(1)$ can
be achieved, therefore the coincidence problem can be solved. The interaction between dark sectors
changes the perturbation dynamics and modifies the cosmic microwave background spectrum \cite{cmbpert}. For the
dark energy model with constant equation of motion parameter $w$, it was found that the curvature perturbation
has a super-Hubble instability in the early radiation dominated era, whenever a particular interaction
term is present \cite{pert1}. A more careful analysis finds that the stability of the curvature perturbation
depends on the form of the interaction between dark sectors \cite{pert2}. The dynamical quintessence
model considered in this Letter may not suffer the instability problem.

By introducing the interaction between dark matter and
dark energy, the conservation equations become
\be
\label{rmcons}
\dot\rho_m+3H\rho_m=-Q,
\ee
\be
\label{rxcons}
\dot\rho_\phi+3H(1+w_\phi)\rho_\phi=Q,
\ee
where the dark matter energy density is $\rho_m$, the dark energy density
$\rho_\phi=\dot{\phi}^2/2+V(\phi)$, the dark energy pressure $p_\phi=\dot{\phi}^2/2-V(\phi)$,
the equation of state of the dark energy $w_\phi=p_\phi/\rho_\phi$ and $Q$ stands for the interaction term.
The phenomenological interaction term $Q$ is inspired from the interaction between the dilaton field $\sigma$ and the matter field
in the scalar-tensor theory of gravity \cite{kaloper},
\be
\label{bdaction}
{\cal L}= \sqrt{-g} \left[-\frac{1}{2\kappa^2}{\mathcal
R} -\frac{1}{2}g^{\mu\nu}\partial_\mu \sigma \partial_\nu
\sigma -\xi(\sigma)^{-2}{\cal L}_{m}(\psi, \xi(\sigma)^{-1}g_{\mu\nu})\right].
\ee
For a general coupling function $\xi(\sigma)$, the interaction term $Q=-3\rho_m H [d(\ln\xi)/d(\ln a)]/2$ \cite{quiros}.

The Letter is organized as follows. In section 2, we review the method of the phase-plane
analysis by studying the model discussed in \cite{bohmer}. In section 3, we discuss
the interaction model $Q=\alpha_0 \kappa^2 H^{-1} \rho_m^2$ and its accelerated scaling attractors.
In section 4, we study the interaction model $Q=\beta \kappa^2 H^{-1} \rho_m \dot\phi^2$
and its accelerated scaling attractors. We conclude the Letter in section 5.

\section{Interacting model 1}
In this section, we consider the interaction $Q=\alpha H \rho_m$ \cite{bohmer} to show
the phase-plane analysis. Using the dimensionless variables
\be
\label{xydef}
x^2=\frac{\kappa^2\dot{\phi}^2}{6H^2},\quad y^2=\frac{\kappa^2 V}{3H^2},
\ee
Eqs. (\ref{rmcons}), (\ref{rxcons}) and the Friedmann equation become
\be
\label{mod1xeq1}
x'=-3x+\frac{\sqrt{6}}{2}\lambda y^2+\frac{3}{2}x(1+x^2-y^2)+\alpha\frac{1-x^2-y^2}{2x},
\ee
\be
\label{mod1yeq1}
y'=-\frac{\sqrt{6}}{2}\lambda xy+\frac{3}{2}y(1+x^2-y^2),
\ee
where a prime denotes $d/d\ln a$. Setting $x'=0$ and $y'=0$, we find that
the fixed points of the autonomous system (\ref{mod1xeq1}) and (\ref{mod1yeq1}) are
\be
\label{mod1fxpts1}
\begin{split}
(x_{c1}=\pm 1,\ y_{c1}=0),\quad (x_{c2}=\pm \sqrt{\frac{\alpha}{3}},\ y_{c1}=0),
\quad (x_{c3}=\frac{\lambda}{\sqrt{6}},\ y_{c3}=\sqrt{1-\frac{\lambda^2}{6}}),\\
(x_{c4}=\frac{\alpha+3}{\sqrt{6}\lambda},\ y_{c4}=\frac{\sqrt{(\alpha+3)^2-2\alpha\lambda^2}}{\sqrt{6}\lambda}).\quad \quad
\end{split}
\ee

\begin{figure}[htp]
\centering
\includegraphics[width=14cm]{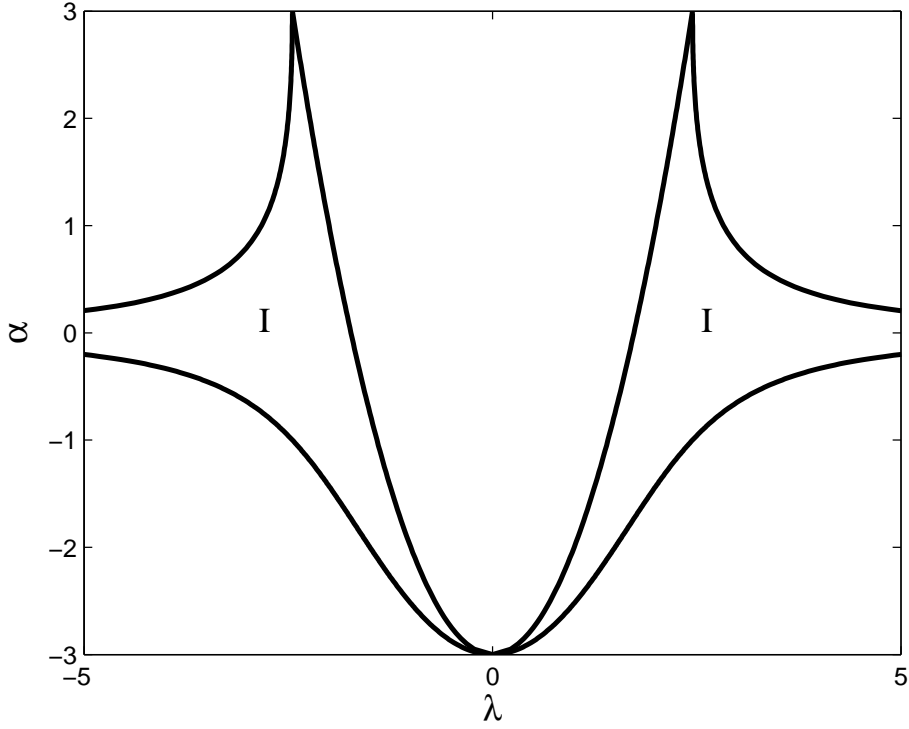}
\caption{The region I is the parameter space for the fixed point ($x_{c4}$, $y_{c4}$) in model 1 to be stable.}
\label{mod1par}
\end{figure}

\begin{figure}[htp]
\centering
\includegraphics[width=14cm]{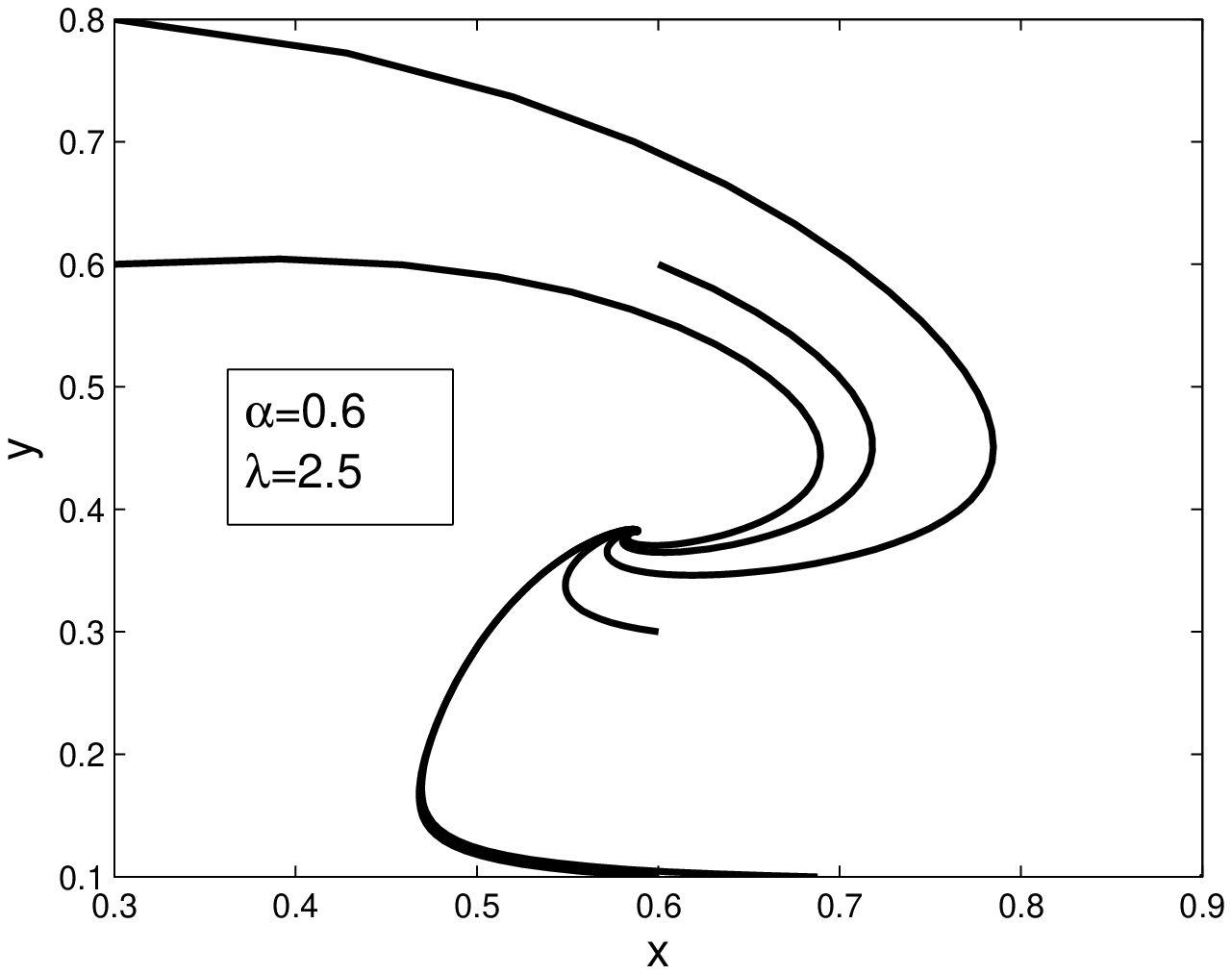}
\caption{Phase-space trajectories for the model 1 with $\alpha=0.6$ and $\lambda=2.5$, the stable fixed point
is ($x_{c4}$, $y_{c4}$)=(0.59, 0.38).}
\label{cont1a}
\end{figure}

\begin{figure}[htp]
\centering
\includegraphics[width=14cm]{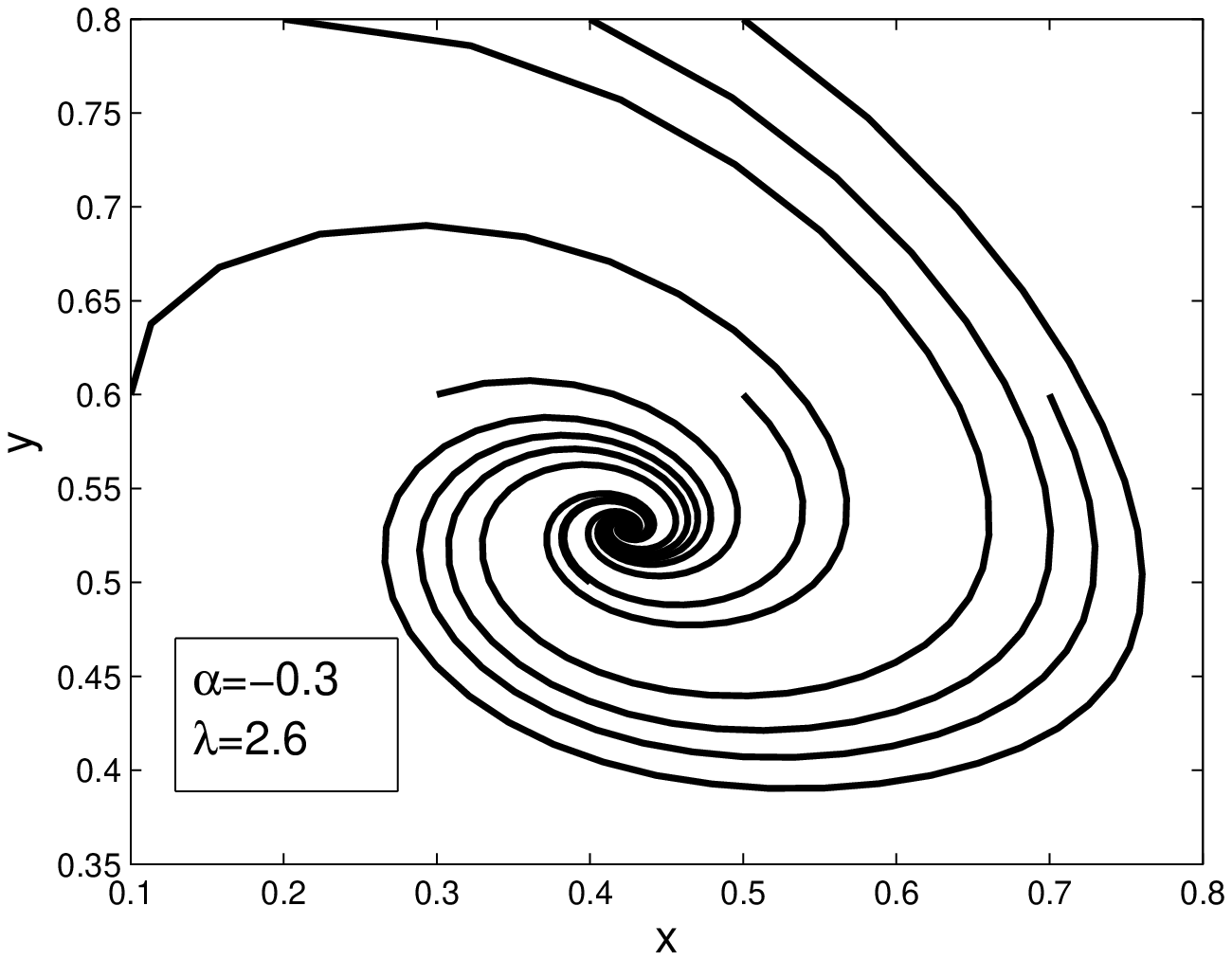}
\caption{Phase-space trajectories for the model 1 with $\alpha=-0.3$ and $\lambda=2.6$,
the stable fixed point
is ($x_{c4}$, $y_{c4}$)=(0.42, 0.53).}
\label{cont1b}
\end{figure}

In terms of the variables $x$ and $y$, the dark energy density and the equation of state of the total matter are
\be
\label{pardef}
\Omega_\phi=x^2+y^2,\quad w_{tot}=\frac{p_\phi}{\rho_\phi+\rho_m}=x^2-y^2.
\ee
Since the model was already discussed in \cite{bohmer} and the other fixed points are not interesting for our purpose,
here we use the fixed point ($x_{c4}$, $y_{c4}$) as an example to discuss
the stability of the fixed point. For the existence of the the fixed point ($x_{c4}$, $y_{c4}$), we require
$(\alpha+3)^2-2\alpha\lambda^2\ge 0$ and $0\le \Omega_\phi\le 1$. Therefore, we get the existence conditions
\begin{eqnarray}
\lambda^2\ge \alpha+3, \quad & 0\ge \alpha \ge -3 \nonumber \\
\frac{(\alpha+3)^2}{2\alpha} \ge \lambda^2  \ge \alpha+3, \quad& 3 \ge \alpha > 0, \label{mod1exs1}
\end{eqnarray}
To discuss the stability of the fixed point, we need to expand the system (\ref{mod1xeq1}) and (\ref{mod1yeq1}) around the fixed point.
In general, for an autonomous system
\be
\label{genxeq}
x'=f(x,y), \quad y'=g(x,y),
\ee
we have a constant nonsingular matrix at the fixed point ($x_c$, $y_c$),
\be
\label{matrxdef}
M=\begin{pmatrix}
a_{11}=\frac{\partial f}{\partial x}(x_c,\ y_c)& a_{12}=\frac{\partial f}{\partial y}(x_c,\ y_c)\\
a_{21}=\frac{\partial g}{\partial x}(x_c,\ y_c)& a_{22}=\frac{\partial g}{\partial y}(x_c,\ y_c)
\end{pmatrix}.
\ee
The eigenvalues of the matrix $M$ are
\be
\label{meigen}
\frac{a_{11}+a_{22}\pm \sqrt{(a_{11}+a_{22})^2-4(a_{11}a_{22}-a_{12}a_{21})}}{2}.
\ee
If the real parts of the eigenvalues of the matrix $M$ are negative, then the fixed point is a stable point. So the conditions for the
fixed point to be stable are
\be
\label{stblecond}
a_{11}+a_{22}<0,\quad a_{11}a_{22}-a_{12}a_{21}>0.
\ee
Combining equations (\ref{mod1xeq1}), (\ref{mod1yeq1}) and the conditions (\ref{stblecond}), we find that the stability conditions
for the fixed point ($x_{c4}$, $y_{c4}$) are
\begin{eqnarray}
\frac{3(\alpha+3)(\alpha-1)}{2\alpha}>\lambda^2 > \alpha+3, \quad & 0\ge \alpha \ge -3 \nonumber \\
\frac{(\alpha+3)^2}{2\alpha} \ge \lambda^2  > \alpha+3, \quad& 3 \ge \alpha > 0. \label{mod1stb1}
\end{eqnarray}
The result is plotted in Fig.
\ref{mod1par}. From Fig. \ref{mod1par}, we see that the parameter space for the fixed point to be stable is much larger
than that obtained in \cite{bohmer}. To verify the correctness of our result, we numerically solve
the system equations (\ref{mod1xeq1}) and (\ref{mod1yeq1}) with different initial conditions for the
parameters ($\alpha$, $\lambda$)=(0.6, 2.5) and ($\alpha$, $\lambda$)=(-0.3, 2.6). The results are shown in Figs.
\ref{cont1a} and \ref{cont1b}. The parameters are outside the parameter space for the
fixed point ($x_{c4}$, $y_{c4}$) to be stable in \cite{bohmer}, but satisfy our conditions (\ref{mod1exs1}).
From Figs. \ref{mod1par}, \ref{cont1a} and \ref{cont1b},
we see that the fixed point
($x_{c4}$, $y_{c4}$) is a stable point when ($\alpha$, $\lambda$)=(0.6, 2.5) or ($\alpha$, $\lambda$)=(-0.3, 2.6).

\section{Interacting model 2}

The interaction $Q=\alpha H \rho_m$ can take the more general form $Q=\alpha_0 \kappa^{2n-2} H^{3-2n} \rho_m^n$,
so that the dynamical system still has the attractor solution $x^2+y^2=1$. In this section, we consider
$n=2$ for simplicity. The autonomous system is
\be
\label{mod2xeq1}
x'=-3x+\frac{\sqrt{6}}{2}\lambda y^2+\frac{3}{2}x(1+x^2-y^2)+\frac{3}{2}\alpha_0\frac{(1-x^2-y^2)^2}{x},
\ee
\be
\label{mod2yeq1}
y'=-\frac{\sqrt{6}}{2}\lambda xy+\frac{3}{2}y(1+x^2-y^2).
\ee
For this system, the fixed points are
\be
\label{mod2fxpts1}
\begin{split}
(x_{c1}=\pm 1,\ y_{c1}=0),\quad (x_{c2}=\pm \sqrt{\frac{\alpha_0}{1+\alpha_0}},\ y_{c2}=0),
\quad (x_{c3}=\frac{\lambda}{\sqrt{6}},\ y_{c3}=\sqrt{1-\frac{\lambda^2}{6}}),\\
(x_{c4},\ y_{c4}=\sqrt{1+x^2_{c4}-\sqrt{6}\lambda x_{c4}/3}),\quad
(x_{c5},\ y_{c5}=\sqrt{1+x^2_{c5}-\sqrt{6}\lambda x_{c5}/3}),
\end{split}
\ee
where
\be
\label{xc4def}
x_{c4}=\frac{(\alpha_0-1)\lambda + \sqrt{\lambda^2(\alpha_0-1)^2+12\alpha_0}}{2\sqrt{6}\alpha_0},
\ee
and
\be
\label{xc5def}
x_{c5}=\frac{(\alpha_0-1)\lambda - \sqrt{\lambda^2(\alpha_0-1)^2+12\alpha_0}}{2\sqrt{6}\alpha_0}.
\ee

The accelerated attractors that solve the coincidence problem are the fixed points ($x_{c4}$, $y_{c4}$)
and ($x_{c5}$, $y_{c5}$). For the existence of the fixed point ($x_{c4}$, $y_{c4}$), we require
\begin{eqnarray}
\sqrt{3}\le \lambda \le \sqrt{\frac{3(1+2\alpha_0)^2}{2\alpha_0(1+\alpha_0)}}=\lambda_u, & \alpha_0> 0 \nonumber \\
\lambda  \ge \sqrt{3}, \quad& -1\le \alpha_0\le 0, \label{mod2exs1}\\
\lambda \ge \sqrt{\frac{-12\alpha_0}{(\alpha_0-1)^2}}=\lambda_l\ {\rm or}\ -\sqrt{3}\le \lambda \le -\lambda_l,\quad
& \alpha_0<-1. \nonumber
\end{eqnarray}

\begin{figure}[htp]
\centering
\includegraphics[width=14cm]{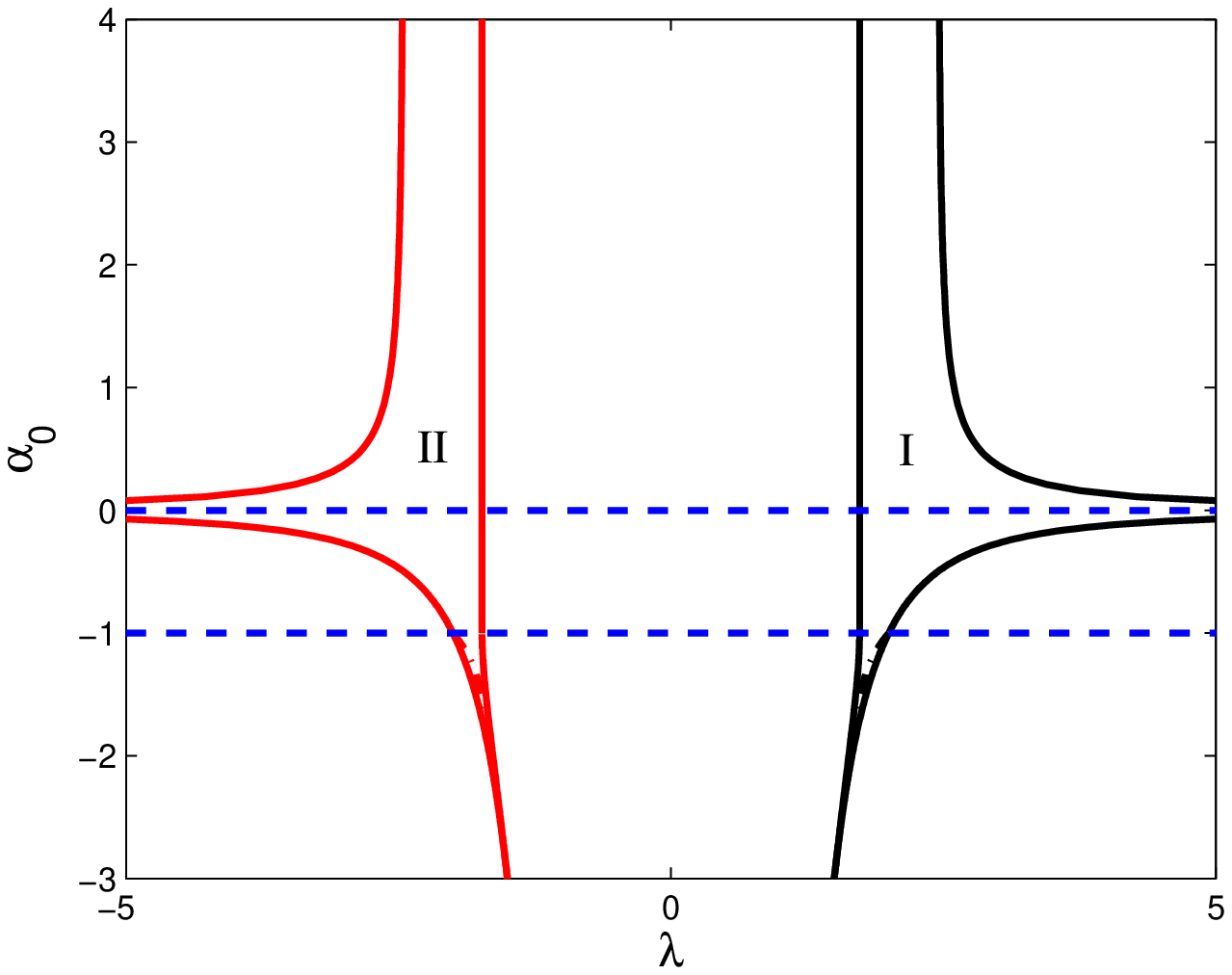}
\caption{The stability conditions for the fixed points in model 2.
The region I is the parameter space
for the fixed point ($x_{c4}$, $y_{c4}$) to be stable and the region II is the parameter space
for the fixed point ($x_{c5}$, $y_{c5}$) to be stable. The dash-dot line denotes
the acceleration condition.}
\label{mod2par}
\end{figure}

The stability conditions are
\begin{eqnarray}
\sqrt{3}<\lambda \le \lambda_u, & \alpha_0>0,\nonumber\\
\sqrt{3}<\lambda < \lambda_+, \quad & -1/2\le \alpha_0 < 0, \nonumber\\
\sqrt{3}<\lambda < \lambda_{-}, \quad & -1< \alpha_0< -1/2, \nonumber\\
\lambda_l \le \lambda < \lambda_{-}, \quad & -3 <\alpha_0 <-1, \label{mod2stb1}
\end{eqnarray}
where
\be
\label{lambda1}
\lambda_+=\sqrt{\frac{-3(1-2\alpha_0+\sqrt{1-4\alpha_0-28\alpha_0^2
-32\alpha_0^3}\,)}{4\alpha_0(1+\alpha_0)}},
\ee
and
\be
\label{lambda2}
\lambda_{-}=\sqrt{\frac{-3(1-2\alpha_0-\sqrt{1-4\alpha_0-28\alpha_0^2
-32\alpha_0^3}\,)}{4\alpha_0(1+\alpha_0)}}.
\ee
The regions of the parameters $\alpha_0$ and $\lambda$ for the fixed point ($x_{c4}$, $y_{c4}$) to be stale
are plotted in the region I in Fig. \ref{mod2par}.

Note that the stability condition for the fixed point ($x_{c3}$, $y_{c3}$) is $\lambda^2\le 3$. From Fig. \ref{mod2par}, we
see that for $\alpha_0<-1$ and $\lambda<\sqrt{3}$, both the fixed points ($x_{c3}$, $y_{c3}$) and ($x_{c4}$, $y_{c4}$) are
stable points, so they are local stable points for those parameters. In other words, when $\alpha_0<-1$ and $\lambda<\sqrt{3}$,
different initial conditions may lead to either the fixed point ($x_{c3}$, $y_{c3}$) or ($x_{c4}$, $y_{c4}$).

To get acceleration, we require $w_{tot}=x^2-y^2<-1/3$. For the stable point ($x_{c4}$, $y_{c4}$),
the acceleration conditions are
\begin{eqnarray}
\lambda>\sqrt{\frac{4\alpha_0}{1+2\alpha_0}}, \quad &  -2<\alpha_0<-1 \nonumber\\
\lambda\ge \lambda_l, \quad &  \alpha\le-2. \label{mod2acc1}
\end{eqnarray}
The accelerated region is also shown in Fig. \ref{mod2par}.

For the fixed point ($x_{c5}$, $y_{c5}$), the existence conditions and the stability conditions
can be obtained from those of the fixed point ($x_{c4}$, $y_{c4}$) by replacing $\lambda$
with $-\lambda$ in equations (\ref{mod2exs1}) and (\ref{mod2stb1}). The existence conditions are
\begin{eqnarray}
-\sqrt{3} \ge \lambda \ge -\lambda_u, \quad& \alpha_0> 0 \nonumber \\
\lambda \le -\sqrt{3}, \quad& -1\le \alpha_0\le 0, \label{mod2exs2}\\
\sqrt{3} \ge \lambda \ge \lambda_l\ {\rm or}\ \lambda \le -\lambda_l ,\quad
& \alpha_0<-1. \nonumber
\end{eqnarray}

The stability conditions are
\begin{eqnarray}
-\sqrt{3} > \lambda \ge -\lambda_u, & \alpha_0>0,\nonumber\\
-\sqrt{3} > \lambda > -\lambda_+, \quad & -1/2\le \alpha_0 < 0, \nonumber\\
-\sqrt{3} > \lambda > -\lambda_{-}, \quad & -1< \alpha_0< -1/2, \nonumber\\
-\lambda_l \ge \lambda > -\lambda_{-}, \quad & -3 <\alpha_0 <-1, \label{mod2stb2}
\end{eqnarray}
The condition (\ref{mod2stb2}) is shown in the region II in Fig. \ref{mod2par}.
The acceleration conditions are
\begin{eqnarray}
\lambda<-\sqrt{\frac{4\alpha_0}{1+2\alpha_0}}, \quad & -2<\alpha_0<-1, \nonumber\\
\lambda\le -\lambda_l, \quad & \alpha\le-2.\label{mod2acc2}
\end{eqnarray}
These results are summarized in Table \ref{tab1}.
\begin{table}[htp]
\begin{tabular}{|l|c|c|c|c|c|c|c|} \hline
$x$ & $y$ & Stability Condition & $\Omega_\phi$ & Acceleration Condition \\\hline
1 & 0 & Unstable & 1 & No \\\hline
-1 & 0 & Unstable & 1 & No \\\hline
$\sqrt{\frac{\alpha_0}{1+\alpha_0}}$ & 0 & $\alpha_0>0$, $\lambda>\lambda_u$  & $\frac{\alpha_0}{1+\alpha_0}$ & No \\\hline
$-\sqrt{\frac{\alpha_0}{1+\alpha_0}}$ & 0 & $\alpha_0>0$, $\lambda<-\lambda_u$  & $\frac{\alpha_0}{1+\alpha_0}$ & No \\\hline
$\lambda/\sqrt{6}$ & $\sqrt{1-\lambda^2/6}$ & $\lambda^2<3$ & 1 & $\lambda^2<2$ \\\hline
$x_{c4}$ & $y_{c4}$ & Equation (\ref{mod2stb1}) & $x_{c4}^2+y_{c4}^2$ & Equation (\ref{mod2acc1}) \\\hline
$x_{c5}$ & $y_{c5}$ & Equation (\ref{mod2stb2}) & $x_{c5}^2+y_{c5}^2$ & Equation (\ref{mod2acc2}) \\\hline
\end{tabular}
 \caption{The behaviors of the fixed points in model 2.} \label{tab1}
\end{table}

From Fig. \ref{mod2par}, we see that the attractors ($x_{c4}$, $y_{c4}$)
and ($x_{c5}$, $y_{c5}$) lead to accelerated scaling attractors only when $-\lambda_{-}<\lambda<-(4\alpha_0/(1+2\alpha_0))^{1/2}$ or
$\lambda_{-}>\lambda>(4\alpha_0/(1+2\alpha_0))^{1/2}$
if $-2<\alpha_0<-1$ and $-\lambda_l\ge \lambda >-\lambda_{-}$ or $\lambda_l\le \lambda <\lambda_{-}$ if $-3<\alpha_0\le -2$.

\section{Interacting Model 3}
In this section, we take the interaction term $Q=\beta \kappa^{2n} H^{1-2n}\rho_m^n \dot\phi^2$. The dynamical
system has the attractors $x^2+y^2=1$.
For simplicity, we consider $n=1$ case, the autonomous system is
\be
\label{mod3xeq1}
x'=-3x+\frac{\sqrt{6}}{2}\lambda y^2+\frac{3}{2}x(1+x^2-y^2)+3\beta x(1-x^2-y^2),
\ee
\be
\label{mod3yeq1}
y'=-\frac{\sqrt{6}}{2}\lambda xy+\frac{3}{2}y(1+x^2-y^2).
\ee
The fixed points are
\be
\label{mod3fxpts1}
\begin{split}
(x_{c1}=\pm 1,\ y_{c1}=0),\quad (x_{c2}=0,\ y_{c2}=0),
\quad (x_{c3}=\frac{\lambda}{\sqrt{6}},\ y_{c3}=\sqrt{1-\frac{\lambda^2}{6}}),\\
(x_{c4},\ y_{c4}=\sqrt{1+x^2_{c4}-\sqrt{6}\lambda x_{c4}/3}),\quad
(x_{c5},\ y_{c5}=\sqrt{1+x^2_{c5}-\sqrt{6}\lambda x_{c5}/3}),
\end{split}
\ee
where
\be
\label{mod3xc4}
x_{c4}=\frac{\lambda+\sqrt{\lambda^2-12\beta}}{2\sqrt{6}\beta},
\ee
and
\be
\label{mod3xc4}
x_{c5}=\frac{\lambda-\sqrt{\lambda^2-12\beta}}{2\sqrt{6}\beta}.
\ee

\begin{figure}[htp]
\centering
\includegraphics[width=14cm]{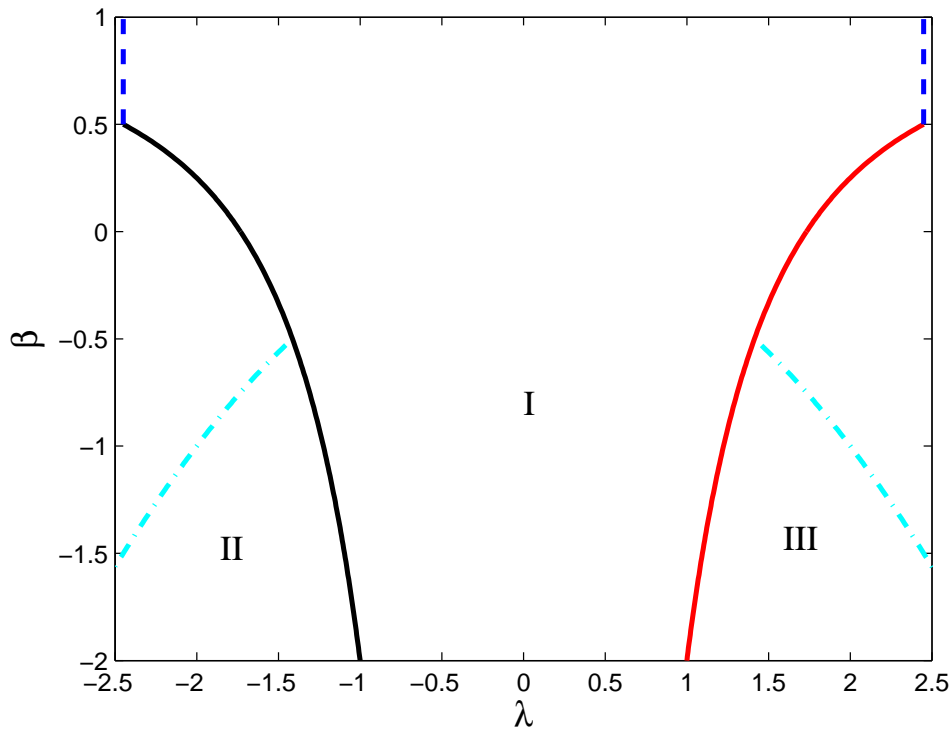}
\caption{The stability conditions for the fixed points ($x_{c3}$, $y_{c3}$), ($x_{c4}$, $y_{c4}$)
and ($x_{c5}$, $y_{c5}$) in model 3. The region I is the parameter space
for the fixed point ($x_{c3}$, $y_{c3}$) to be stable, the region II is the parameter space
for the fixed point ($x_{c4}$, $y_{c4}$) to be stable and the region III is the parameter space
for the fixed point ($x_{c5}$, $y_{c5}$) to be stable. The dash-dot line denotes
the acceleration condition.}
\label{mod3par}
\end{figure}

For the fixed point ($x_{c3}$, $y_{c3}$), the existence condition is
$\lambda^2\le 6$, and the stability conditions are
\begin{eqnarray}
-\sqrt{6}\le \lambda \le \sqrt{6},\quad & \beta\ge 1/2, \nonumber\\
-\sqrt{\frac{3}{1-\beta}}\ < \lambda < \sqrt{\frac{3}{1-\beta}},\quad & \beta<1/2.\label{mod3stab3}
\end{eqnarray}
The condition (\ref{mod3stab3}) is shown in the region I in Fig. \ref{mod3par}.
The acceleration condition is $-\sqrt{2} < \lambda < \sqrt{2}$.

For the fixed point ($x_{c4}$, $y_{c4}$), the existence condition is
\be
\label{mod3exs1}
\lambda \le -\sqrt{\frac{3}{1-\beta}}, \quad \beta\le \frac{1}{2},
\ee
and the stability condition is
\be
\label{mod3stab1}
\lambda < -\sqrt{\frac{3}{1-\beta}},\quad \beta\le 1/2.
\ee
the condition (\ref{mod3stab1}) is shown in the region II in Fig. \ref{mod3par}.
The acceleration condition is $-\sqrt{3/(1-\beta)} \ge \lambda > -\sqrt{-4\beta}$ and $\beta<0$.

For the fixed point ($x_{c5}$, $y_{c5}$), the existence condition is
\be
\label{mod3exs2}
\lambda \ge \sqrt{\frac{3}{1-\beta}}, \quad \beta\le \frac{1}{2},
\ee
and the stability conditions are
\be
\label{mod3stab2}
\lambda > \sqrt{\frac{3}{1-\beta}},\quad \beta\le 1/2.
\ee
The condition (\ref{mod3stab2}) is shown in the region III in Fig. \ref{mod3par}.
The acceleration condition is $\sqrt{3/(1-\beta)} \le \lambda < \sqrt{-4\beta}$ and $\beta<0$.
These results are summarized in Table \ref{tab3}.

\begin{table}[htp]
\begin{tabular}{|l|c|c|c|c|c|c|c|} \hline
$x$ & $y$ & Stability Condition & $\Omega_\phi$ & Acceleration Condition \\\hline
1 & 0 & $\beta>1/2$ and $\lambda>\sqrt{6}$ & 1 & No \\\hline
-1 & 0 & $\beta>1/2$ and $\lambda<-\sqrt{6}$ & 1 & No \\\hline
0 & 0 & Unstable & 0 & No \\\hline
$\lambda/\sqrt{6}$ & $\sqrt{1-\lambda^2/6}$ & Equation (\ref{mod3stab3}) & 1 & $\lambda^2<2$ \\\hline
$x_{c4}$ & $y_{c4}$ & Equation (\ref{mod3stab1}) & $x_{c4}^2+y_{c4}^2$ & $\lambda>-\sqrt{-4\beta}$ and $\beta<0$ \\\hline
$x_{c5}$ & $y_{c5}$ & Equation (\ref{mod3stab2}) & $x_{c5}^2+y_{c5}^2$ & $\lambda<\sqrt{-4\beta}$ and $\beta<0$ \\\hline
\end{tabular}
 \caption{The behaviors of the fixed points in model 3.} \label{tab3}
\end{table}

\section{Discussions}

We considered two phenomenological interacting models $Q=\alpha_0 H^{3-2n}\rho_m^n$ and $Q=\beta H^{1-2n}\rho_m^n\dot\phi^2$.
The scaling attractor solutions $x^2+y^2=1$ for the non-interacting quintessence model with exponential potential
remain to be the scaling attractors in the interacting models. We have studied the dynamical behaviors of
the two interacting dark energy models.  For the interacting
model $Q=\alpha_0 H^{-1}\rho_m^2$, we find that the fixed points ($x_{c3}$, $y_{c3}$) and ($x_{c4}$, $y_{c4}$)
are stable points if $\alpha_0<-1$ and $\lambda<\sqrt{3}$, and the fixed points
($x_{c3}$, $y_{c3}$) and ($x_{c5}$, $y_{c5}$)
are stable points if $\alpha_0<-1$ and $\lambda>-\sqrt{3}$. In other words, in
the parameter region $\alpha_0<-1$ and $\lambda<\sqrt{3}$ or $\lambda>-\sqrt{3}$,
the fixed points ($x_{c3}$, $y_{c3}$), ($x_{c4}$, $y_{c4}$) and ($x_{c5}$, $y_{c5}$)
are local stable points. This type of local stable points are new in the dark energy models.

These models have the late time accelerated scaling attractors
with $\Omega_\phi/\Omega_m=O(1)$. We can easily match $\Omega_\phi/\Omega_m$ to observations by a simple
choice of parameters. Since the solution is a scaling attractor, the value of $\Omega_\phi/\Omega_m$
is insensitive to initial conditions, therefore the why now problem is resolved. For
the interacting model $Q=\beta H^{-1} \rho_m\dot\phi^2$, when we choose $\beta=-2.0$
and $\lambda=2.5$, the stable fixed point is ($x_{c5}$, $y_{c5}$)=(0.31, 0.68) with
$w_{tot}=-0.37$ and $\Omega_\phi=0.75$. Note that to get accelerated attractor solution
and alleviate the coincidence problem, we find that $\alpha<0$ and $\beta<0$, so
the energy transfer goes from dark energy to dark matter. This result is easily understood.
The energy transfer from dark energy to dark matter makes the dark matter to decrease slower and
dark energy to decrease faster, therefore alleviating the coincidence problem.

\begin{acknowledgments}
The work is supported by NNSFC under grant No. 10605042. The authors would like to thank E.N. Saridakis
for valuable comments.
\end{acknowledgments}


\begin{thebibliography}{sp99}
\bibitem{quint} B. Ratra, P.J.E. Peebles, Phys. Rev. D 37 (1988) 3406; C. Wetterich, Nucl. Phys. B 302 (1988) 668;
R.R. Caldwell, R. Dave, P.J. Steinhardt, Phys. Rev. Lett. 80 (1998) 1582;
I. Zlatev, L. Wang, P.J. Steinhardt, Phys. Rev. Lett. 82 (1999) 896.
\bibitem{holo} M. Li, Phys. Lett. B 603 (2004) 1;
Y.G. Gong, Phys. Rev. D 70 (2004) 064029;
B. Wang, Y.G. Gong, E. Abdalla, Phys. Lett. B 624 (2005) 141;
M.R. Setare,  Phys. Lett. B 642 (2006) 1;
M.R. Setare, E.C. Vagenas, Int. J. Mod. Phys. D 18 (2009) 147;
M.R. Setare, Phys. Lett. B 654 (2007) 1;
Y.G. Gong, J. Liu, JCAP 0809 (2008) 010.
\bibitem{chaplyg} A. Yu. Kamenshchik, U. Moschella, V. Pasquier, Phys. Lett. B 511 (2001) 265;
N. Bilic, G.B. Tupper, R.D. Viollier, Phys. Lett. B 535 (2002) 17;
M.C. Bento, O. Betrolami, A.A. Sen, Phys. Rev. D 66 (2002) 043507;
Y.G. Gong, JCAP 0503 (2005) 007.
\bibitem{tach} A. Mazumdar, S. Panda, A. P\'{e}rez-Lorenzana, Nucl. Phys. B 614 (2001) 101;
A. Sen, JHEP 0207 (2002) 065;
T. Padmanabhan, Phys. Rev. D 66 (2002) 021301;
A. Feinstein, Phys. Rev. D 66 (2002) 063511;
T.R. Choudhury, T. Padmanabhan, Phys. Rev. D 66 (2002) 081301.
\bibitem{weak} Q.G. Huang, JHEP 0705 (2007) 096;
Q.G. Huang, Phys. Rev. D 77 (2008) 103518;
X. Wu, Z.H. Zhu, Chin. Phys. Lett. 25 (2008) 1517;
X.M. Chen, J. Liu, Y.G. Gong, Chin. Phys. Lett. 25 (2008) 3086.
\bibitem{invr} S.M. Carroll, V. Duvvuri, M. Trodden, M.S. Turner, Phys. Rev. D 70 (2004) 043528;
T. Chiba, Phys. Lett. B 575 (2003) 1;
S. Nojiri, S.D. Odintsov, Phys. Rev. D 68 (2003) 123512;
C.G. Shao, R.G. Cai, B. Wang, R.K. Su, Phys. Lett. B 633 (2006) 164.
\bibitem{fr} S. Capozziello, Int. J. Mod. Phys. D 11 (2002) 483;
S. Nojiri, S.D. Odintsov, Int. J. Geom. Meth. Mod. Phys. 4 (2007) 115;
W. Hu, I. Sawicki, Phys. Rev. D 76 (2007) 064004;
S. Nojiri, S.D. Odintsov, arXiv: 0807.0685.
\bibitem{dgp} G.R. Dvali, G. Gabadadze, M. Porrati, Phys. Lett. B 485 (2000) 208;
C. Deffayet, G.R. Dvali, G. Gabadadze, Phys. Rev. D 65 (2002) 044023;
Y.G. Gong, C.K. Duan, Class. Quantum Grav. 21 (2004) 3655;
Y.G. Gong, C.K. Duan, Mon. Not. Roy. Astron. Soc. 352 (2004) 847;
Y.G. Gong, Phys. Rev. D 78 (2008) 123010.
\bibitem{brane} P. Bin\'{e}truy, C. Deffayet, D. Langlois, Nucl. Phys. B 565 (2000) 269;
R.G. Cai, Y.G. Gong, B. Wang, JCAP 0603 (2006) 006;
Y.G. Gong, A. Wang, Class. Quantum Grav. 23 (2006) 3419;
Y.G. Gong, A. Wang, Q. Wu, Phys. Lett. B 663 (2008) 147.
\bibitem{expon} P.G. Ferreira, M. Joyce, Phys. Rev. Lett. 79 (1997) 4740;
E.J. Copeland, A.R. Liddle, D. Wands, Phys. Rev. D 57 (1998) 4686.
\bibitem{gong} E.J. Copeland, M. Sami, S. Tsujikawa, Int. J. Mod. Phys. D 15 (2006) 1753;
Y.G. Gong, A. Wang, Y.Z. Zhang, Phys. Lett. B 636 (2006) 286.
\bibitem{intmod} A. Nunes, J.P. Mimoso, T.C. Charters, Phys. Rev. D 63 (2001) 083506;
W. Zimdahl, D. Pav\'{o}n and L.P. Chimento, Phys. Lett. B 521 (2001) 133;
L.P. Chimento, A.S. Jakubi, D. Pav\'{o}n and W. Zimdahl, Phys. Rev. D 67 (2003) 083513;
D.F. Mota, C. van de Bruck, Astron. Astrophys. 421 (2004) 71;
M. Manera, D.F. Mota, Mon. Not. Roy. Astron. Soc. 371 (2006) 1373;
N.J. Nunes, D.F. Mota, Mon. Not. Roy. Astron. Soc. 368 (2006) 751;
J.D. Barrow, T. Clifton, Phys. Rev. D 73 (2006) 103520;
T. Clifton, J.D. Barrow, Phys. Rev. D 73 (2006) 104022;
T. Clifton, J.D. Barrow, Phys. Rev. D 75 (2007) 043515;
M.R. Setare, E.N. Saridakis, JCAP 0809 (2008) 026;
M.R. Setare, E.N. Saridakis, Phys. Lett. B 668 (2008) 177;
S.~del Campo, R. Herrera and D. Pav\'{o}n, Phys. Rev.  D 78 (2008) 021302;
T. Gonzalez and I. Quiros, Class. Quantum Grav.  25 (2008) 175019;
M. Jamil, M.A. Rashid, Eur. Phys. J. C 60 (2009) 141;
M. Jamil, arXiv: 0810.2896;
S.~del Campo, R. Herrera and D. Pav\'{o}n, arXiv:0812.2210 [gr-qc];
M.R. Setare, E.N. Saridakis, arXiv: 0810.4775.
\bibitem{cmbpert} B. Wang etal., Nucl. Phys. B 778 (2007) 69;
G. Olivares, F. Atrio-Barandela and D. Pav\'{o}n, Phys. Rev. D 77 (2008) 103520.
\bibitem{pert1} J. V\"{a}liviita, E. Majerotto and R. Maartens, JCAP 0807 (2008) 020.
\bibitem{pert2} J.-H. He, B. Wang and E. Abdalla, Phys. Lett. B 671 (2009) 139.
\bibitem{kaloper} N. Kaloper and K.A. Olive, Phys. Rev. D 57 (1998) 811.
\bibitem{quiros} R. Curbelo, T. Gonzalez, G. Leon and I. Quiros, Class. Quantum Grav. 23 (2006) 1585.
\bibitem{bohmer} C.G. B\"{o}hmer, G. Caldera-Cabral, R. Lazkoz, R. Maartens, Phys. Rev. D 78 (2008) 023505.

\end{thebibliography}
\end{document}